\documentclass[
   prd,
   amsfonts,
   amssymb,
   amsmath,
   superscriptaddress,
   twocolumn,
   eqsecnum,
   showpacs,
   preprintnumbers,
   ]{revtex4}
\usepackage{color}
\usepackage{graphicx}
\usepackage{subfigure}
\usepackage{time}
\newcommand{\oneplusone}{(1+1)}            
\newcommand{\threeplusone}{(3+1)}          
\newcommand{\QED}{\textit{QED}}            
\newcommand{\QCD}{\textit{QCD}}            
\newcommand{\QFT}{\textit{QFT}}            
\newcommand{\QGP}{\textit{QGP}}            
\newcommand{\RHIC}{\textit{RHIC}}          
\newcommand{\LHC}{\textit{LHC}}            
\newcommand{\BV}{\textit{B-\!V}}           
\newcommand{\SUtwo}{\textit{SU(2)}}        
\newcommand{\SUthree}{\textit{SU(3)}}      
\newcommand{\SUN}{\textit{SU(N)}}          

\newcommand{\etal}{\textit{et~al.}}        
\newcommand{\rd}{\mathrm{d}}               
\newcommand{\rD}{\mathrm{D}}               
\newcommand{\coD}{\mathcal{D}}             
\newcommand{\BVop}{\mathcal{B}}            
\newcommand{\defby}{\equiv}                
\newcommand{\kt}{k_{t}}                    
\newcommand{\kz}{k_{z}}                    
\newcommand{\pz}{p_{z}}                    
\newcommand{\kperp}{k_{\perp}}             
\newcommand{\pperp}{p_{\perp}}             
\newcommand{\bA}{\mathbf{A}}               
\newcommand{\bD}{\mathbf{D}}               
\newcommand{\bE}{\mathbf{E}}               
\newcommand{\bF}{\mathbf{F}}               
\newcommand{\bJ}{\mathbf{J}}               
\newcommand{\bQ}{\mathbf{Q}}               
\newcommand{\bT}{\mathbf{T}}               
\newcommand{\bV}{\mathbf{V}}               
\newcommand{\Eeigen}{\tilde{E}}            
\newcommand{\Evec}{e}                      
\newcommand{\Diag}[1]{\mathrm{diag}( \, #1 \, )}   
\newcommand{\Set}[1]{ \bigl ( \, #1 \, \bigr )}    
\newcommand{\Expect}[1]
   {\ensuremath{\langle \, #1 \,  \rangle}}
\newcommand{\Sgn}[1]{\mathrm{sgn}[ \, #1 \, ]}     
\newcommand{\Comm}[2]
   {\ensuremath{[ \, #1, #2 \, ]}}
\newcommand{\AntiComm}[2]
   {\ensuremath{\{ \, #1, #2 \, \}}}
\newcommand{\PB}[2]
   {\ensuremath{\{ \, #1, #2 \, \}}}
\newcommand{\Tr}[1]{\mathrm{Tr} [ \, #1 \, ]}      
\begin{document}
%
%
%
\begin{figure}[!t]
\vskip -1.3cm
\leftline{ \includegraphics[width=1.1cm]{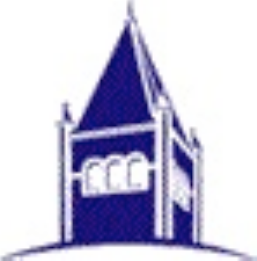} }
\end{figure}
\preprint{\vbox{
   \hbox{LA-UR-10-00512}
   \hbox{UNH-Theory-10-1} } }
\title[BV-QCD]
   {Dynamics of particle production by strong electric fields 
     in non-Abelian plasmas}

\author{John F. Dawson}
\email{john.dawson@unh.edu}
\affiliation{Department of Physics,
   University of New Hampshire,
   Durham, NH 03824}

\author{Bogdan Mihaila}
\email{bmihaila@lanl.gov}
\affiliation{Materials Science and Technology Division,
   Los Alamos National Laboratory,
   Los Alamos, NM 87545}

\author{Fred Cooper}
\email{cooper@santafe.edu}
\affiliation{National Science Foundation,
   4201 Wilson Blvd.,
   Arlington, VA 22230}
\affiliation{Santa Fe Institute,
   Santa Fe, NM 87501}
\affiliation{Center for Nonlinear Studies,
   Los Alamos National Laboratory,
   Los Alamos, NM 87545}



\begin{abstract}
We develop methods for computing the dynamics of fermion pair production by strong color electric fields using the semi-classical Boltzmann-Vlasov equation.  We present numerical results for a model with \SUtwo\ symmetry in \oneplusone\ Cartesian dimensions. 
\end{abstract}
\maketitle

%
%
\section{Introduction}
\label{s:intro}

High energy collisions of heavy ions at \RHIC, and soon to be carried out at the \LHC, are thought to produce a quark-gluon plasma (\QGP), yet efforts to model this plasma using quantum field theory methods have proved to be particularly difficult.  Since heavy ion collisions are dynamic events, finding appropriate approximation schemes are important.  The \QCD-\SUthree\ case, for example, requires the use of three quark colors for each flavor and eight gluon fields.  The non-Abelian character of \QCD, preserving gauge invariance, and dynamic renormalization schemes all add to the difficulties.  

In a recent paper, we have demonstrated that classical theories based on Boltzmann-Vlasov (\BV) transport equations have worked well for the \QED\ problem in \oneplusone\ and \threeplusone\ dimensions in both Cartesian and boost-invariant coordinates.  So it seems reasonable to try to develop a transport equation for non-Abelian gauge theories.  Fortunately, such transport equations have been worked out for some time, using the Wong equations \cite{r:Wong:1970yq}.  (See, for example, Refs.~\onlinecite{r:Litim:2002kx} and \onlinecite{r:Schenke:2008ve} and references therein.)  In this paper, we apply \BV\ transport equation methods to find the classical distribution function for particles with \SUtwo\ internal symmetry with an initial large electric field in the $z$-direction and no particles present at $t=0$.  Particles are created by means of the Schwinger mechanism for particles with internal symmetry.  We derive the necessary equations in \threeplusone\ dimensions in Cartesian coordinates, but only give numerical results in \oneplusone\ dimensions.  The methods used here can be extended to boost-invariant coordinates in \threeplusone\ dimensions and \SUthree\ symmetry.  

The Wong equations have also been used to develop a theory of non-Abelian fluid dynamics.  See Refs.~\onlinecite{r:Bistrovic:2003mz} and \onlinecite{r:Jackiw:2003gf}, and to study unstable quark-gluon plasmas \cite{r:Manuel:2006ly}.

Similar work for the \SUtwo\ case using a Wigner function has been carried out by Skokov and L{\'e}vai \cite{r:Skokov:2008kx} using methods described in Prozorkevich, \etal~\cite{r:Prozorkevich:2004vn}.

We discuss the \QFT\ problem in Section~\ref{s:backreaction}.  In Section~\ref{s:classpareqs} we introduce the Wong equations for a particle with non-Abelian internal symmetry and find conserved quantities for the case we consider.  In Section~\ref{s:distfct}, we define particle distribution functions and in Section~\ref{s:BVequ} we write down the \BV\ equation for our case.  In Section~\ref{s:creation}, we give the Schwinger formulas for particle creation, modified for our non-Abelian case, and in Section~\ref{s:BVsolution}, we find a solution of the \BV\ equation.  In Section~\ref{s:polcurrents} we derive polarization currents.  Numerical results are given in Section~\ref{s:numerical}, and conclusions in Section~\ref{s:conclusions}.

%
%
\section{The back-reaction problem in QCD}
\label{s:backreaction}

The colored quarks obey the Dirac equation,
\begin{equation}\label{e:DiracI}
   \bigl \{ \,
      \gamma^{\mu} \,
      [ \, i \partial_{\mu} + g \, A_{\mu}(x) \, ]
      -
      M \,
   \big \} \, \psi(x) 
   =
   0 \>,
\end{equation}
where $A_{\mu}(x) = A^{a}_{\mu}(x) \, T^{a}$.  The generators obey a Lie algebra: $\Comm{T^a}{T^b} = i f^{abc} T^c$.  The sign of $g$ here is chosen to correspond to that in Ref.~\onlinecite{r:Litim:2002kx}.  In this paper we consider the situation in Cartesian coordinates where $g_{\mu\nu} = \Diag{1,-1,-1,-1}$.  From Dirac's Eq.~\eqref{e:DiracI}, the color current,
\begin{equation}\label{e:colorCurrentDirac}
   J^{a,\mu}(x)
   =
   g \, 
   \Expect{ \bar{\psi}(x) \, \gamma^{\mu} \, T^a \, \psi(x) } \>, 
\end{equation}
obeys a conservation relation, $D^{a,c}_{\mu}(x) \, J^{c,\mu}(x)=0$, where
\begin{equation}\label{e:DiracColorCons}
   D^{a,c}_{\mu}(x)
   \defby
   \delta^{a,c} \partial_{\mu}
   +
   g \, f^{abc} \, A_{\mu}^b(x) \>.
\end{equation}
For the gauge field, the gauge field tensor $F_{\mu\nu}(x) = F_{\mu\nu}^a(x) \, T^a$ is given by
\begin{equation}\label{e:FdefII}
   F_{\mu\nu}^a(x)
   =
   \partial_{\mu}^{\phantom{a}} A_{\nu}^a(x)
   -
   \partial_{\nu}^{\phantom{a}} A_{\mu}^a(x)
   + 
   g \, f^{abc} \,
   A_{\mu}^b(x) A_{\nu}^c(x) \>.
\end{equation}
and satisfies the equation of motion,
\begin{equation}\label{e:FeomII}
   D^{a,c}_{\mu}(x) \, F^{c,\mu\nu}(x)
   =
   \Expect{ J^{a,\nu} } \>.
\end{equation}
In this last equation, we have taken the expectation value of the Dirac color current in some initial state of the system.  The gauge field is treated as a classical commuting field.  Eqs.~\eqref{e:DiracI} and \eqref{e:FeomII} constitute the \QCD-\QFT\ problem with feedback.

In order to simplify this problem, we choose a gauge and require that the field $A^{a,\mu}(x)$ be in the $z$-direction and depend only on $t$ for all $a$.  That is, we put
\begin{equation}\label{e:Arequire}
   A^{a,\mu}(x)
   =
   \Set{ 0,0,0,A^a(t) } \>.
\end{equation}
Then, from \eqref{e:FdefII}, the only non-vanishing terms are given by
\begin{equation}\label{e:Fmunuvalues}
   F_{t,z}^a(t)
   =
   -
   F_{z,t}^a(t)
   =
   -
   \partial_t A^a(t)
   \defby
   E^a(t) \>.
\end{equation}
From \eqref{e:FeomII}, for $\mu=t$ and $\nu=z$, we find
\begin{equation}\label{e:caseI}
   \partial_t \, F^{a,tz}(t)
   =
   - \partial_t \, E^{a}(t)
   =
   J^{a,z}(t) \>.
\end{equation}
We can also have $\mu=z$ and $\nu=t$, in which case we find
\begin{equation}\label{e:caseII}
   g \, f^{abc} A^b_z(t) \, F^{c,zt}(t)
   =
   g \, f^{abc} A^b(t) \, E^c(t)
   =
   J^{a,t}(t) \>.
\end{equation}
We can write these field equations in an obvious vector notation as
\begin{subequations}\label{e:fieldeqs}
\begin{align}
   \partial_{t} \, \bA(t)
   &=
   - \bE(t) \>,
   \label{e:fielda} \\
   \partial_{t} \, \bE(t)
   &=
   - \bJ^z(t) \>,
   \label{e:fieldb} \\
   g \,
   \bA(t) \times \bE(t)
   &=
   \bJ^t(t) \>,
   \label{e:fieldc} 
\end{align}
\end{subequations}  
Taking the derivative of \eqref{e:fieldc} with respect to $t$ and using \eqref{e:fieldb} gives a conservation equation:
\begin{equation}\label{e:Jtcons}
   \partial_t \bJ^t(t)
   +
   g \, \bA(t) \times \bJ^z(t) 
   =
   0 \>.
\end{equation}

The gauge field energy-momentum tensor is given by
\begin{equation*}
   \Theta^{\mu\nu}(x)
   =
   \frac{1}{4} \, 
   g^{\mu\nu} \,
   F_{\alpha\beta}^a(x) \, F^{a,\alpha\beta}(x)
   +
   F^{a,\mu\alpha}(x) \, g_{\alpha\beta} \, F^{a,\beta\nu}(x) \>,
\end{equation*}
and using the field equations \eqref{e:FeomII}, the divergence of the field tensor reads
\begin{equation}\label{e:fieldengmomcons}
   \partial_{\mu} \, \Theta^{\mu\nu}(x)
   =
   - \bF^{\nu\sigma}(x) \cdot \bJ_{\sigma}(x) \>.
\end{equation}
For the case we consider here, the field tensor can be written as
\begin{equation}\label{e:fielddiag}
   \Theta^{\mu\nu}(t) 
   = 
   \frac{E^2(t)}{2} \, 
   \Diag{ 1,1,1,-1} \>.
\end{equation}
The conservation equation \eqref{e:fieldengmomcons} reduces to the two equations
\begin{subequations}\label{e:twoenergyeqs}
\begin{align}
   \partial_{t} \, [ \, E^2(t) / 2 \, ]
   &=
   \bE(t) \cdot [ \, \partial_t \bE(t) \, ]
   =
   - \bE(t) \cdot \bJ^z(t) \>,
   \label{e:energya} \\
   0
   &=
   \bE(t) \cdot \bJ^t(t) \>,
   \label{e:energyb}    
\end{align}
\end{subequations}
both of which are a result of dotting $\bE(t)$ into the equations of motion \eqref{e:fieldeqs}.

%
%
\section{Classical particle equations}
\label{s:classpareqs}

A classical microscopic theory of colored charges interacting with a non-Abelian gauge field has been developed by Wong in 1970 \cite{r:Wong:1970yq}.  We follow the development in Litim and Manuel~\cite{r:Litim:2002kx}. The Wong theory for \SUN\ assigns to each particle on a trajectory $x^{\mu}(s)$ in space-time $N^2-1$ color charges $Q^a(s)$ which depend on the trajectory proper time.  The color charges are to be thought of as replacing average values of the group generators, $Q^a(s) \mapsto \Expect{ T^a(s) }$, in the Heisenberg representation.

The proper time interval $\rd s$ is defined by
\begin{equation}\label{e:dsdef}
   ( \rd s )^2
   =
   g_{\mu\nu}(x) \, \rd x^{\mu} \rd x^{\nu} \>,
\end{equation}
and the four velocity $u^{\mu}(s)$ and kinetic momentum $k^{\mu}(s)$ along the path is defined by
\begin{align}
   k^{\mu}(s) 
   &\defby 
   M \, u^{\mu}(s)
   \defby
   M \, \frac{\rd x^{\mu}(s)}{\rd s}
   \label{e:ukmomdef} \\
   &\defby
   \Set{\kt(s),k_x(s),k_y(s),\kz(s)} \>.
   \notag
\end{align}
Instead of starting with a Lagrangian, we simply state the Wong equations of motion for non-Abelian fields,
\begin{subequations}\label{e:eomI}
\begin{align}
   M \, \frac{\rd k_{\mu}(s)}{\rd s}
   &=
   g \, F_{\mu\nu}^a(s) \, k^{\nu}(s) \, Q^a(s) \>,
   \label{e:eomIa} \\
   M \, \frac{\rd Q^a(s)}{\rd s}
   &=
   - g \, f^{abc} \, A_{\mu}^b(s) \, Q^c(s) \, k^{\mu}(s) \>.
   \label{e:eomIb}
\end{align}
\end{subequations}
Writing $Q^a(s)$ as a function of $x^{\mu}(s)$, we have
\begin{equation}\label{e:notethat}
   \frac{\rd Q^a(s)}{\rd s}
   =
   u^{\mu}(s) \,
   \frac{\partial Q^a(x)}{\partial x^{\mu}} \>,
\end{equation}
so that the equation of motion for $Q^a(s)$, Eq.~\eqref{e:eomIb}, can be written in terms of the covariant derivative,
\begin{equation}\label{e:wongd}
   k^{\mu} \, 
   D_{\mu}^{ac}(x) \, Q^{c}(x) 
   =
   0 \>,
\end{equation}
where $D_{\mu}^{ac}(x)$ is given in \eqref{e:DiracColorCons}.  

In a vector notation, Eqs.~\eqref{e:eomI} can be written as
\begin{subequations}\label{e:eomII}
\begin{align}
   M \, \frac{\rd k_{\mu}(s)}{\rd s}
   &=
   g \, \bQ(s) \cdot \bF_{\mu\nu}(s) \, k^{\nu}(s) \>,
   \label{e:eomIIa} \\
   M \, \frac{\rd \bQ(s)}{\rd s}
   &=
   - g \, \bA_{\mu}(s) \times \bQ(s) \, k^{\mu}(s) \>,
   \label{e:eomIIb}   
\end{align}
\end{subequations}
and \eqref{e:wongd} becomes
\begin{equation}\label{e:wongdvec}
   k^{\mu} \, \bD_{\mu}(x) \, \bQ(x) 
   =
   0 \>,
\end{equation}
where $\bD_{\mu}(x)$ is the covariant vector derivative operator
\begin{equation}\label{e:Dvecdef}
   \bD_{\mu}(x)
   =
   \partial_{\mu}
   +
   g \, \bA_{\mu}(x) \times \>.
\end{equation}
Eq.~\eqref{e:eomIIb} says that the color charge  vector $\bQ(s)$ precesses about the vector $\bA_{\mu}(s) \, u^{\mu}(s)$ along the trajectory in space-time.  Multiplying \eqref{e:eomIIa} by $k^{\mu}(s)$, and requiring the momentum to be on the mass shell, leads to the fact that the length of the kinetic momentum vector is conserved, and given by
\begin{equation}\label{e:kcons}
   k_{\mu}(s) \, k^{\mu}(s)
   =
   \kt^2(s) - k_x^2(s) - k_y^2(s) - \kz^2(s)
   =
   M^2 \>,
\end{equation}
or
\begin{equation}\label{e:kts}
   \kt(s)
   =
   \sqrt{ \kperp^2(s) + \kz^2(s) + M^2 } \>,
\end{equation}
where $\kperp^2(s) = k_x^2(s) + k_y^2(s)$.  Dotting $\bQ(s)$ into Eq.~\eqref{e:eomIIb} leads to the observation that the length $Q^2$ of the color charge vector, which is the quadratic Casimir, is conserved.  For $SU(3)$, the cubic Casimir is also conserved.  

For our case, Eq.~\eqref{e:eomIIa} becomes
\begin{subequations}\label{e:eomIII}
\begin{align}
   M \,
   \frac{\rd \kt(s)}{\rd s}
   &=
   g \, \bQ(s) \cdot \bE(s) \, \kz(s) \>,
   \label{e:eomIIIa} \\
   M \, \frac{\rd \kz(s)}{\rd s}
   &=
   g \, \bQ(s) \cdot \bE(s)  \, \kt(s) \>,
   \label{e:eomIIIb}
\end{align}
\end{subequations}
with $k_x$ and $k_y$ (and consequently $\kperp$) constants of the motion.
For our case, Eq.~\eqref{e:eomIIb} becomes
\begin{equation}\label{e:eomIVb}
   M \,
   \frac{\rd \bQ(s)}{\rd s}
   =
   g \, \bA(s) \times \bQ(s) \, \kz(s) \>.
\end{equation}
The dot product of $\bA(s)$ with Eq.~\eqref{e:eomIVb} gives
\begin{equation}\label{e:AadQa}
   M \,
   \bA(s) \cdot \frac{\rd \bQ(s)}{\rd s}
   =
   0 \>,
\end{equation}
and since
\begin{equation}\label{e:AsEs}
   \kt(s) \, \bE(s)
   =
   - M \, \frac{\rd t}{\rd s} \, 
   \frac{\rd \bA(t)}{\rd t}
   =
   - M \,
   \frac{\rd \bA(s)}{\rd s} \>,
\end{equation}
Eq.~\eqref{e:eomIIIb} can be written as a total derivative:
\begin{equation}\label{e:consmoI}
   M \, \frac{\rd}{\rd s} \,
   \bigl [ \, \kz(s) + g \, \bQ(s) \cdot \bA(s) \, \bigr ]
   =
   0 \>.
\end{equation}
So the quantity in brackets is a constant of the motion, which we call $\pz$.  Then we have
\begin{equation}\label{e:pconstdef}
   \kz(s)
   =
   \pz - g \, \bQ(s) \cdot \bA(s) \>.
\end{equation}

For a single particle, the color current $\bJ^{\mu}(x)$ is given by
\begin{equation}\label{e:Jcolordef}
   \bJ^{\mu}(x)
   =
   g \int \rd s \, 
   \bQ(s) \, 
   \frac{\rd x^{\mu}(s)}{\rd s} \,
   \delta^4[ \, x - x(s) \, ] \>,
\end{equation}
and the matter energy momentum tensor $t^{\mu\nu}(x)$ by
\begin{equation}\label{e:matem}
   t^{\mu\nu}(x)
   =
   \int \rd s \,
   \frac{ \rd x^{\mu}(s)}{\rd s} \, k^{\nu}(s) \,
   \delta^4[ \, x - x(s) \, ] \>.
\end{equation}
In the next section, we introduce a distribution function for an ensemble of particles.

%
%
\section{Distribution function}
\label{s:distfct}

We define a particle distribution function $f(x,k,Q)$ such that the average convective particle color current density $\bJ_{\text{con}}^{\mu}(s)$ is given by
\begin{equation}\label{e:Ncurrent}
   \bJ_{\text{con}}^{\mu}(x)
   =
   g \int \rD k \, \rD Q \, k^{\mu} \, \bQ \, f(x,p,Q) \>.
\end{equation}
Similarly, the particle energy-momentum tensor density is given by
\begin{equation}\label{e:tmunudef}
   t^{\mu\nu}(x)
   =
   \int \rD k \, \rD Q \, k^{\mu} \, k^{\nu} \, f(x,p,Q) \>.
\end{equation}
Here the momentum measure $\rD k$ is given by
\begin{equation}\label{e:Dkdef}
   \rD k
   =
   \frac{2r \, \Theta(\kt) \, \delta(k^2 - M^2) \, \rd^4 k }
        { (2\pi)^3 \, \sqrt{-g}} \>,
\end{equation}
where $r$ is a degeneracy factor which counts the number of species.  For one flavor of fermions and anti-fermions with no spin $r=2$, so in \oneplusone\ Cartesian dimensions
\begin{equation}\label{e:Dkonepone}
   \rD k
   =
   \frac{\rd \kz }
        { \pi \, \omega_{\kz} } \>,
   \qquad
   \omega_{\kz}
   =
   \sqrt{ \kz^2 + M^2 } \>.
\end{equation}
In \threeplusone\ dimensions for one flavor of fermions and anti-fermions with spin $r=4$, and this factor is given by
\begin{equation}\label{e:Dkthreepone}
   \rD k
   =
   \frac{ \kperp \rd\kperp \, \rd \kz }
        { \pi^2 \, \omega_{\kperp, \kz} } \>,
   \quad
   \omega_{\kperp,\kz}
   =
   \sqrt{ \kperp^2 + \kz^2 + M^2 } \>.
\end{equation}
In subsequent sections of this paper, we work out the necessary equations in \threeplusone\ dimensions --- the translation to \oneplusone\ dimensions essentially means that we omit the $\kperp$ integration, set $\kperp = 0$, and multiply the currents by $\pi$.  For $SU(2)$, the color measure is
\begin{equation}\label{e:DmeasureSU2}
   \rD Q
   =
   c_R \, \rd^3 Q \,
   \delta( \bQ^2 - q_2 ) 
\end{equation}
where the delta-function expresses conservation of the quadratic Casimir.  
Here $c_R$ is a normalization factor.  $c_R$ and $q_2$ are set by the conditions,
\begin{equation}\label{e:conditions}
\begin{split}
   \Tr{ 1 }
   =
   2
   \quad &\mapsto \quad
   \int \rD Q = 2 \>,
   \\
   \Tr{ \bT \cdot \bT }
   =
   3/2
   \quad &\mapsto \quad
   \int \rD Q \,
   \bQ \cdot \bQ
   =
   3/2 \>.
\end{split}
\end{equation}
Writing $\bQ$ in spherical coordinates,
\begin{equation}\label{cc.ss.e:SU2QQQ}
\begin{split}
   Q^1
   &=
   J \, \sin \theta \, \cos \phi \>,
   \\
   Q^2
   &=
   J \, \sin \theta \, \sin \phi \>,
   \\
   Q^3
   &=
   J \, \cos \theta \>,
\end{split}
\end{equation}
we find from Eqs.~\eqref{e:conditions}, the results: $c_R = 2 / ( \pi \sqrt{3} )$, $q_2 = 3/4$, and $J = \sqrt{3}/2$.  Similar results are found for \SUthree\, where there are two conserved Casimir invariants \cite{r:Litim:2002kx}.

%
%
\section{Boltzmann-Vlasov equation}
\label{s:BVequ}

The Boltzmann-Vlasov equation can be derived in parametric form by considering the total derivative of a function $f[\,x(s),k(s),Q(s)\,]$.  Using the equations of motion \eqref{e:eomI}, we find
\begin{equation}\label{e:BVeqI}
\begin{split}
   &M \, \frac{\rd f[\,x(s),k(s),Q(s)\,]}{\rd s}
   =
   M \,
   \biggl \{ \,
      \frac{\rd x^{\mu}(s)}{\rd s} \,
      \frac{\partial}{\partial x^{\mu}}
      \\ & \qquad
      +
      \frac{\rd k_{\mu}(s)}{\rd s} \,
      \frac{\partial}{\partial k_{\mu}}
      +
      \frac{\rd Q^a(s)}{\rd s} \,
      \frac{\partial}{\partial Q^a} \,
   \biggr \} \, f(x,k,Q)
   \\ & \qquad
   =
   k^{\mu}(s) \, \BVop_{\mu}[A](s) \, f(x,k,Q) \>,
\end{split}      
\end{equation}
where the \BV\ differential operator $\BVop[A](s)$ is defined by
\begin{align}\label{e:BVopdef}
   \BVop_{\mu}[A](s)
   &\defby
   \coD_{\mu}[A]
   -
   g \, \bQ \cdot \bF_{\mu\nu}(x) \,
   \partial_{k_{\nu}} \>,
   \\
   \coD_{\mu}[A]
   &\defby
   \partial_{\mu}
   -
   g \,
   \bA_{\mu}(x) \cdot \bQ \times \partial_{\bQ} \>.  
\end{align}
Here $\coD_{\mu}[A]$ is a color-covariant derivative operator, invariant under gauge transformations.  So now requiring that $k$ to be on the mass shell and $Q$ satisfy  the Casimir relations, the Boltzmann-Vlasov equation is given by
\begin{equation}\label{e:BVeqII}
   k^{\mu} \, \BVop_{\mu}[A](s) \, f(x,k,Q)
   =
   k^t \, C(x,k,Q) \>,
\end{equation}
where $C(x,k,Q)$ is a source term.  For our case, the distribution function is a function of $t$, $\kperp$, $\kz$, and $\bQ$ only.  Then the Boltzmann-Vlasov equation \eqref{e:BVeqII} becomes
\begin{align}
   &\biggl \{ \,
      \frac{\partial}{\partial t}
      +
      g \, \frac{\kz}{\omega_{\kperp,\kz}} \, 
      f^{abc} \, A^b(t) \, Q^c \, 
      \frac{\partial}{\partial Q^a}
      \label{e:BVeqIII}\\ & \quad
      +
      g \, Q^a \,
      E^a(t) \, \frac{\partial }{\partial \kz} \,
   \biggr \} \, f(t,\kperp,\kz,Q)
   =
   C(t,\kperp,\kz,Q) \>.
   \notag
\end{align}

%
%
\subsection{Conservation of color current}
\label{ss:divcurrent}

We first study the divergences of the convective color current and the particle energy-momentum tensor.  For the convective current, we multiply \eqref{e:BVeqII} by $g \, \bQ$ and integrate over $\rD k$ and $\rD Q$.  This gives
\begin{align}
   &g \int \rD k \int \rD Q \> k^{\mu} \, \bQ \,
   \BVop_{\mu}[A] \,  f(x,k,Q)
   \label{e:colorconsI} \\ & \qquad\qquad
   =
   g \int \rD k \int \rD Q \> \bQ \, k^t \, C(x,k,Q) \>.
   \notag
\end{align}
In the first term on the left-hand side of \eqref{e:colorconsI}, the derivative with respect to $x^{\mu}$ can be factored out of the integral.  The second term gives a factor $g \, \bA_{\mu}(x) \times \bJ^{\mu}_{\text{con}}(x)$ by parts integration, and the third term vanishes by parts integration and the antisymmetry of $\bF_{\mu\nu}$, yielding the equation
\begin{subequations}\label{e:colorconsIIVdef}
\begin{gather}
   \bD_{\mu}(x) \, \bJ^{\mu}_{\text{con}}(x)
   =
   g \, \bV(x) \>,
   \label{e:colorconsII} \\
   \bV(x)
   =
   \int \rD k \, \rD Q \, \omega_{\kperp,\kz} \, \bQ \, C(x,k,Q) \>.
   \label{e:Vdef}
\end{gather}
\end{subequations}
where $\bD_{\mu}(x)$ is the covariant vector derivative operator, defined in Eq.~\eqref{e:Dvecdef}.
For our case, the convective color current four-vector is of the form $\bJ_{\text{con}}^{\mu}(t) = \Set{ \bJ_{\text{con}}^{t}(t),0,0,\bJ_{\text{con}}^{z}(t)}$, where
\begin{subequations}\label{e:Jcontzdefs}
\begin{align}
   \bJ_{\text{con}}^{t}(t)
   &=
   \frac{g}{\pi^2} \int_0^{\infty} \!\!\!\kperp \rd\kperp\!\!
   \int_{-\infty}^{+\infty} \!\!\!\rd \kz\!
   \int \!\rD Q
   \label{e:Jcontdef} \\ & \qquad\qquad \times
   \bQ \,
   f(t,\kperp,\kz,Q) \>,
   \notag \\
   \bJ_{\text{con}}^{z}(t)
   &=
   \frac{g}{\pi^2} \int_0^{\infty} \!\!\!\kperp \rd\kperp\!\! 
   \int_{-\infty}^{+\infty} \!\!\!\rd \kz\!
   \int \!\rD Q
   \label{e:Jconzdef} \\ & \qquad\qquad \times
   \frac{\kz}{\omega_{\kperp,\kz}} \, \bQ \,
   f(t,\kperp,\kz,Q) \>,
   \notag
\end{align}
\end{subequations}
So for our case, Eq.~\eqref{e:colorconsII} becomes
\begin{equation}\label{e:colorconsIII}
   \partial_t \bJ_{\text{con}}^{t}(t)
   +
   g \, \bA(x) \times \bJ^{z}_{\text{con}}(t)
   =
   g \, \bV(t) \>.
\end{equation}
The \emph{complete} current from the gauge field equations must satisfy Eq.~\eqref{e:Jtcons}.  We shall see below how to define a polarization current which, when added to the convective current, will produce a total current which satisfies  Eq.~\eqref{e:Jtcons}.

%
%
\subsection{Conservation of energy and momentum}
\label{ss:divenergy}

To find the conservation law for the particle energy-momentum tensor, we multiply \eqref{e:BVeqII} by $k^{\nu}$ and integrate.  This gives
\begin{align}
   &\int \rD k \int\rD Q \> k^{\mu} \, k^{\nu} \,
   \BVop_{\mu}[A] \, f(x,k,Q)
   \label{e:engmomconsI} \\ & \qquad
   =
   \int \rD k \int \rD Q \> k^{\nu} \, k^t \, C(x,k,Q) \>.
   \notag
\end{align}
Again the derivative in the first term on the left-hand side comes out of the integral.  By parts integration, the second term vanishes but the third term yields $- g\, \bQ \cdot \bF^{\nu\sigma}(x) \, k_{\sigma}$, so we find
\begin{subequations}\label{e:engmomconsIIGdef}
\begin{align}
   \partial_{\mu} \, t^{\mu\nu}(x)
   &=
   \bF^{\nu\sigma}(x) \cdot \bJ^{\text{con}}_{\sigma}(x)
   +
   G^{\nu}(x) \>,
   \label{e:engmomA} \\
   G^{\nu}(x)
   &=
   \int [ \rd k ] \int \rD Q \, k^{\nu} \, C(x,k,Q) \>.
   \label{e:GdefB}
\end{align}
\end{subequations}
Following a method introduced by Gatoff, Kerman, and Matsui \cite{r:Gatoff:1987fk}, we will find in Section~\ref{s:polcurrents} below that for our case, we can write
\begin{equation}\label{e:Polcurrent}
   G^{\nu}(x)
   =
   \bF^{\nu\sigma}(x) \cdot \bJ^{\text{pol}}_{\sigma}(x) \>,
\end{equation}
where $\bJ^{\text{pol}}_{\sigma}(x)$ is a polarization current.  Defining the total particle current as $\bJ_{\sigma}(x) = \bJ^{\text{con}}_{\sigma}(x) + \bJ^{\text{pol}}_{\sigma}(x)$, Eq.~\eqref{e:engmomconsIIGdef} becomes
\begin{equation}\label{e:engmomconsIII}
   \partial_{\mu} \, t^{\mu\nu}(x)
   =
   \bF^{\nu\sigma}(x) \cdot \bJ_{\sigma}(x) \>.
\end{equation}
The field energy-momentum tensor satisfies Eq.~\eqref{e:fieldengmomcons}, where $\bJ_{\sigma}(x)$ is the \emph{total} current, so the divergence of the sum of the matter and field energy momentum tensors, $T^{\mu\nu}(x) = t^{\mu\nu}(x) + \Theta^{\mu\nu}(x)$ is given by
\begin{equation}\label{e:engmomconsIIV}
   \partial_{\mu} T^{\mu\nu}(x)
   =
   0 \>,
\end{equation}
and is conserved.  For our case, the particle energy-momentum tensor $t^{\mu\nu}(x)$ is diagonal, and given by
\begin{equation}\label{e:tmunudiag}
   t^{\mu\nu}(t)
   =
   \Diag{ \epsilon(t), p_x(t), p_y(t), p_z(t) } \>,
\end{equation}
where the particle energy and pressure densities are given by 
\begin{subequations}\label{e:tensorcomps}
\begin{align}
   \epsilon(t)
   &=
   \frac{g}{\pi^2} \int_0^{\infty} \!\!\!\kperp \rd\kperp\!\!
   \int_{-\infty}^{+\infty} \!\!\!\rd \kz\!
   \int \!\rD Q 
   \label{e:energy} \\ & \qquad\qquad \times
   \omega_{\kperp,\kz} \,
   f(t,\kperp,\kz,Q) \>,
   \notag \\
   \pperp(t)
   &=
   \frac{g}{\pi^2} \int_0^{\infty} \!\!\!\kperp \rd\kperp\!\!
   \int_{-\infty}^{+\infty} \!\!\!\rd \kz\!
   \int \!\rD Q
   \label{e:pperp} \\ & \qquad\qquad \times
   \frac{\kperp^2}{\omega_{\kperp,\kz}} \,
   f(t,\kperp,\kz,Q) \>,
   \notag \\
   \pz(t)
   &=
   \frac{g}{\pi^2} \int_0^{\infty} \!\!\!\kperp \rd\kperp\!\!
   \int_{-\infty}^{+\infty} \!\!\!\rd \kz\!
   \int \!\rD Q
   \label{e:pz} \\ & \qquad\qquad \times
   \frac{\kz^2}{\omega_{\kperp,\kz}} \,
   f(t,\kperp,\kz,Q) \>.
   \notag
\end{align}
\end{subequations}
with $p_x(t) = p_y(t)$ and $\pperp(t) = p_x(t) + p_y(t)$.  

%
%
\section{Particle creation}
\label{s:creation}

%
%
\subsection{\QFT\ pair production}
\label{ss:QFT}

Pair production rates via the Schwinger mechanism by strong non-Abelian gauge fields have recently been calculated using a one-loop approximation in \QFT\ for bosons by Nayak and Nieuwenhuizen \cite{ref:NayakNei} and for fermions by Nayak \cite{r:Nayak:2005ao}.  The key to the calculation was to diagonalize the matrix $E \equiv \sum_a E^a \, T^a$.  Let us write the eigenvalue equation for the matrix $E$ as
\begin{equation}\label{e:eigenQFT}
   E \, \Evec_i
   =
   \Eeigen_i \, \Evec_i \>.
\end{equation}
Then the rate of fermion pair production in \threeplusone\ dimensions is given by a sum over the eigenvalues \cite{r:Nayak:2005ao},
\begin{align}
   &\int_{-\infty}^{+\infty} \rd \kz \,
   \frac{\rd^7 N}
        {\rd t \, \rd^3 x \, \rd^3 k}
   \label{e:rateQFT} \\
   & \qquad =
   -
   \sum_i
   \frac{| g \Eeigen_i|}{4 \pi^3} \,
   \ln
   \Bigl \{ \, 
      1 
      - 
      \exp 
      \Bigl [ \, 
         - 
         \frac{ \pi \, ( \kperp^2 + M^2 )}
              { | \, g \Eeigen_i  \, | } \, 
      \Bigr ] \, 
   \Bigr \} \>,
   \notag
\end{align}
where $M$ is the quark mass.  For the case of \SUtwo, the eigenvalues are given by $\Eeigen_{\pm} = \pm | \bE | / 2$, so the two eigenvalues give identical contributions.    

%
%
\begin{figure}[t!]
   \centering
   \includegraphics[width=\columnwidth]{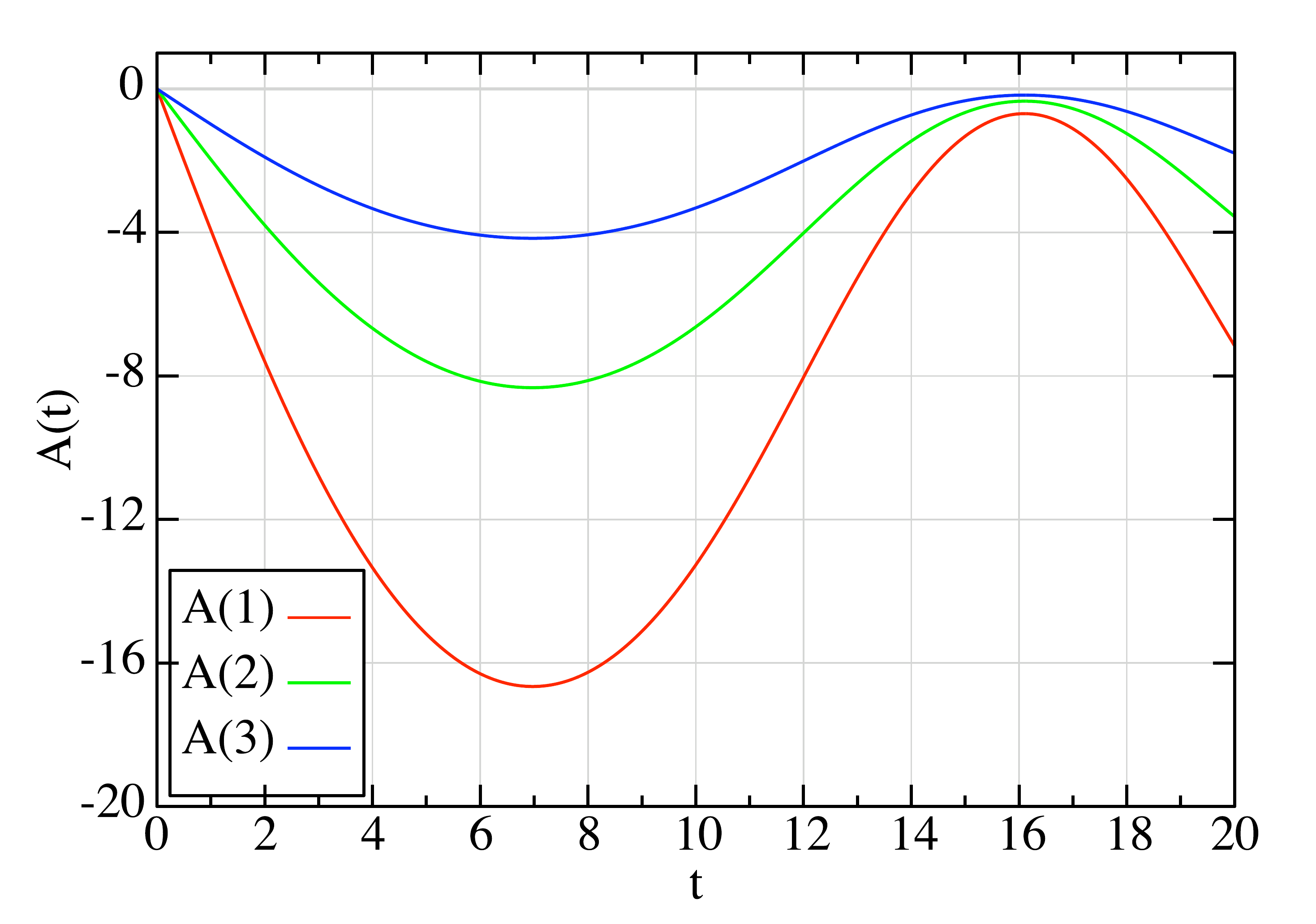}
   \caption{\label{f:A}
   Plot of $\bA(t)$ as a function of $t$ for a solution of the \BV\ equation
   with feedback for $SU(2)$ gauge fields. 
   At $t=0$, we took $\bA(0) = 0$ and $\bE(0)=(4,2,1)$, 
   with $M=1$ and $g=1$.}
\end{figure}
%
%
\begin{figure}[t!]
   \centering
   \includegraphics[width=\columnwidth]{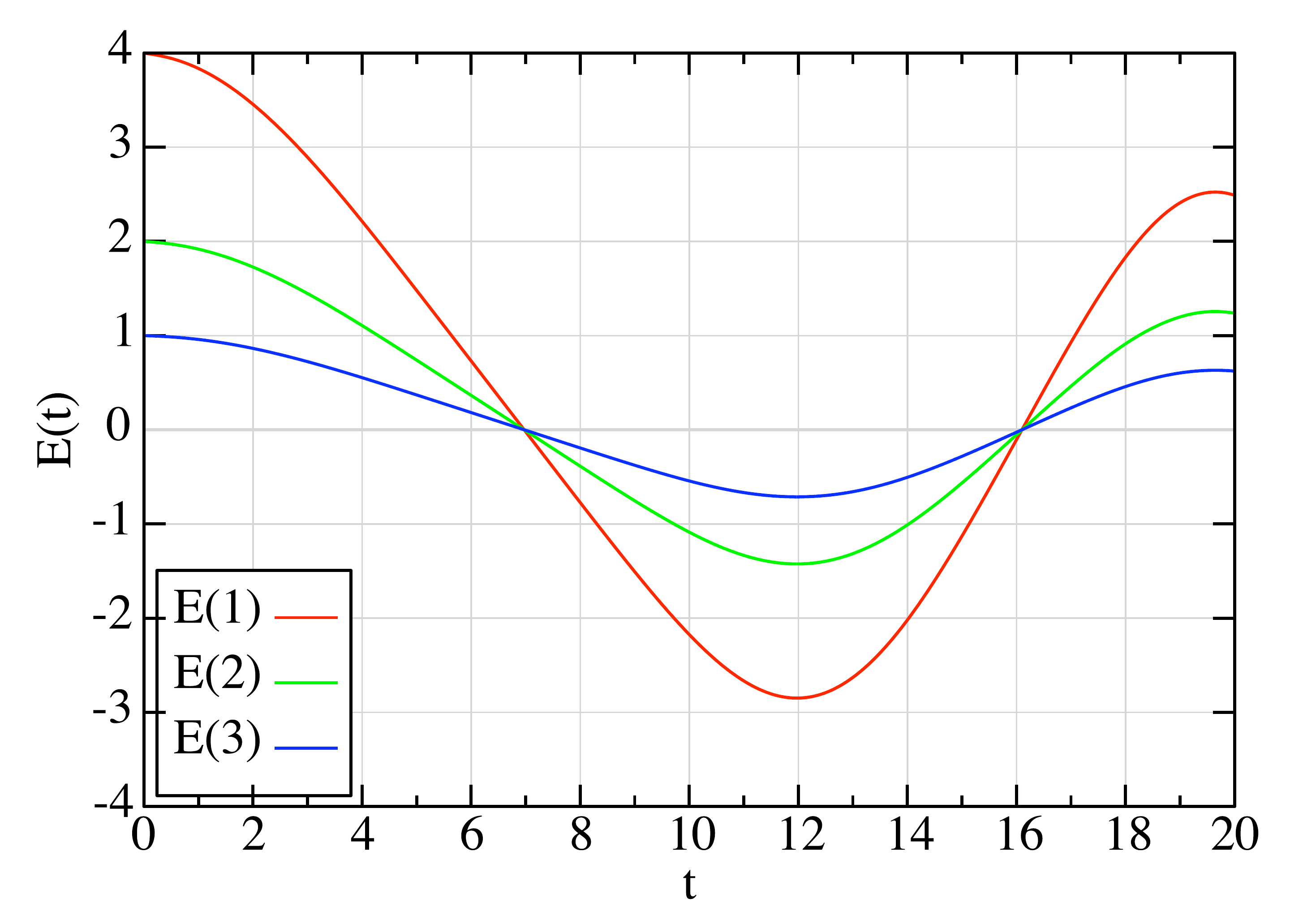}
   \caption{\label{f:E}
   Plot of $\bE(t)$ as a function of $t$ for a solution of the \BV\ equation
   with feedback for $SU(2)$ gauge fields.  Same initial conditions.}
\end{figure}
%
%
%
%
\subsection{Classical pair creation rate}
\label{ss:classcharges}

For the classical calculation, it would seem natural to map $E^a(t) T^a \mapsto \bE(t) \cdot \bQ$, and replace the sum over eigenvalues by an integral over $Q$.
This mapping produces a $Q$-dependent particle production rate $C(t,\kperp,\kz,Q)$ given by
\begin{align}
   C(t,\kperp,\kz,Q)
   &=
   | \, g \, \bQ \cdot \bE(t) \, | \, 
   R(t,\kperp,Q) \,
   \delta(\kz) \>,
   \label{e:stermI} \\
   R(t,\kperp,Q)
   &=
   P(t,\kperp,Q) \, 
   S(t,\kperp,Q) \>,
   \notag
\end{align}
where
\begin{subequations}\label{e:PSdef}
\begin{align}
   P(t,\kperp,Q)
   &=
   1 - 2 f_0(t,\kperp,Q) \>,
   \label{e:Pdefa} \\
   S(t,\kperp,Q)
   &=
   -
   \Bigl \{ \, 
      1 
      - 
      \exp 
      \Bigl [ \, 
         - 
         \frac{ \pi \, ( \kperp^2 + M^2 )}
              { | \, g \, \bQ \cdot \bE(t) \, | } \, 
      \Bigr ] \, 
   \Bigr \} \>,
   \label{e:PSdefb}  
\end{align}
\end{subequations}
and where we have set $f_0(t,\kperp,Q) \equiv f(t,\kperp,0,Q)$.  Here we have multiplied \eqref{e:rateQFT} by $\delta(k_z)$, since most of the particle production occurs at $k_z=0$, and introduced a Pauli supression factor at $k_z = 0$. 
%
%
\begin{figure}[t!]
   \centering
   \includegraphics[width=\columnwidth]{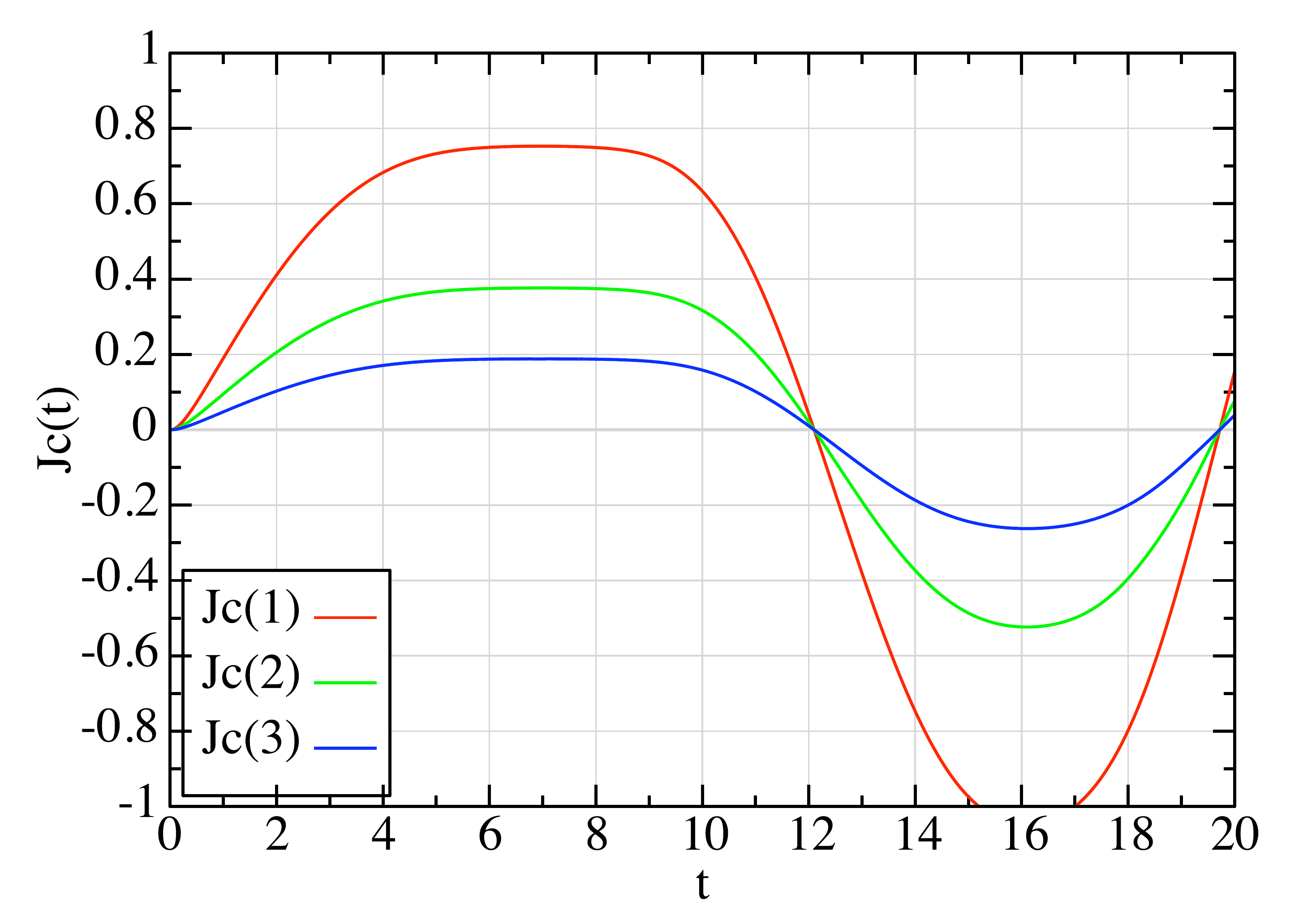}
   \caption{\label{f:Jcz}
   Plot of $\bJ_{\text{con}}^{z}(t)$ as a function 
   of $t$ for a solution of the \BV\ equation
   with feedback for $SU(2)$ gauge fields.  Same initial conditions.}
\end{figure}
%
%
\begin{figure}[t!]
   \centering
   \includegraphics[width=\columnwidth]{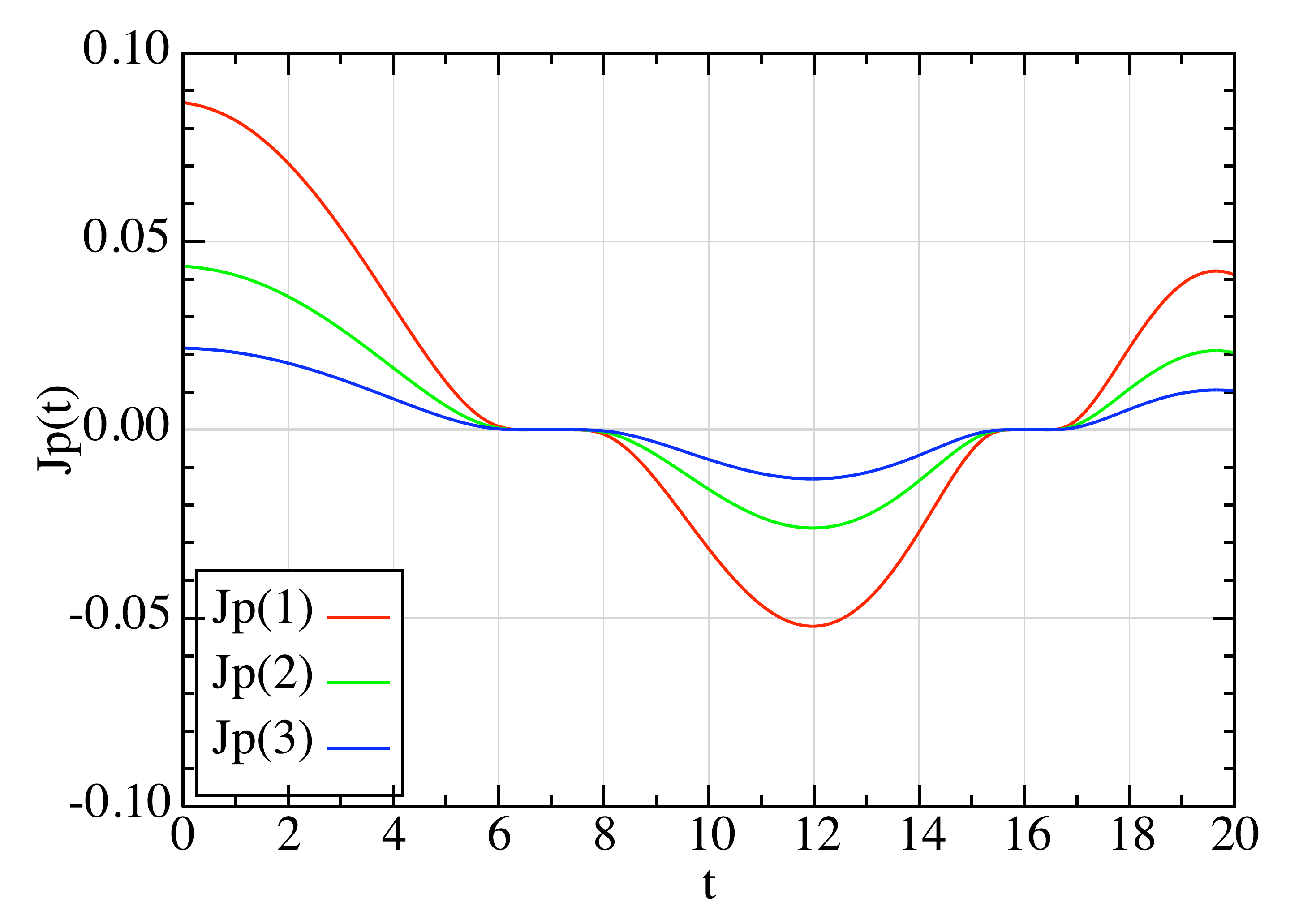}
   \caption{\label{f:Jpz}
   Plot of $\bJ_{\text{pol}}^{z}(t)$ as a function 
   of $t$ for a solution of the \BV\ equation
   with feedback for $SU(2)$ gauge fields.  Same initial conditions.}
\end{figure}
\section{Solution of the Boltzmann-Vlasov equation}
\label{s:BVsolution}

In this section, we solve the BV equation using the method of characteristics.  From Eqs.~\eqref{e:BVeqI} and \eqref{e:BVeqII}, a solution of the BV equation can be found by integrating the source term over a classical particle path trajectory from $s_0$ to $s$,
\begin{align}
   &f[ \, x(s),k(s),Q(s) \, ]
   =
   f[ \, x(s_0),k(s_0),Q(s_0) \, ]
   \label{e:BVsolI} \\
   & \qquad +
   \int_{s_0}^{s} \rd s' \, \kt(s') \, 
   C[ x(s'), k(s'), Q(s') \, ] / M \>,
   \notag
\end{align}
Here $s$ represents the path-length in space-time and $x(s')$, $k(s')$ and $Q(s')$ are solutions of the particle trajectory equations \eqref{e:eomIII} for values of $s$ between $s_0$ and $s$.  The trajectory equations must be integrated \emph{backwards}, starting with the ``current'' values of $x$, $k$, and $Q$, and winding up with a value of $s_0$ such that $t(s_0) = 0$.  We arbitrarily choose $s=0$ so the values of $s'$ are negative.  Then the ``initial'' conditions for the trajectory solutions are
\begin{equation}\label{e:szerochoice}
   x(0) = x \>,
   \qquad
   k(0) = k \>,
   \qquad
   Q(0) = Q \>,
\end{equation}
which are the \emph{current} values of $x$, $k$, and $Q$, and the final value of $s_0$ defined by $t(s_0) = 0$.  Since no particles are present at $t=0$, $f[ \, 0,k(s_0),Q(s_0) \, ] = 0$, and \eqref{e:BVsolI} becomes
\begin{align}
   &f( \, t, \kperp, \kz, Q \, )
   \label{e:BVsolII} \\
   & \qquad =
   \int_{s_0}^{0} \rd s' \, 
   k_t(s') \, C[ \, s', \kperp, \kz(s'), Q(s') \, ] / M
   \notag 
\end{align}
and using \eqref{e:stermI}, we find
\begin{align}
   &f( \, t, \kperp, \kz, Q \, )
   =
   \int_{s_0}^{0} \rd s' \, k_t(s') \, 
   \label{e:BVintIsol} \\ & \qquad \times
   | \, g \bQ(s') \cdot \bE(s') \, | \,  
   R[ \, s',\kperp,Q(s') \, ] \, 
   \delta[ \, \kz(s') \, ] / M \>.
   \notag
\end{align}
From \eqref{e:pconstdef}, we have
\begin{align}
   \kz(s')
   &=
   \pz - g \, \bQ(s') \cdot \bA(s')
   \label{e:kzsp} \\
   &=
   \kz + g \, [ \, \bQ \cdot \bA(s) - \bQ(s') \cdot \bA(s') \, ] \>. 
   \notag
\end{align}
Inserting this into the delta-function in \eqref{e:BVintIsol}, and noting that
\begin{equation}\label{e:dkzdsp}
\begin{split}
   M \, \frac{\rd \kz(s')}{\rd s'}
   &=
   g \, \bQ(s') \cdot \bE(s') \, k_t(s') \>,
\end{split}
\end{equation}
we find that
\begin{align}
   &f( \, t, \kperp, \kz, Q \, )
   \label{e:fsumsnI} \\
   & \qquad =
   \sum_{n}
   R[ \, s_n,\kperp,Q(s_n) \, ] \, 
   \Theta[ \, t(s_n) \, ] \, \Theta[ \, t - t(s_n) \, ] \>,
   \notag
\end{align}
where $s_n$ is a solution of the equation
\begin{equation}\label{e:snsolI}
   \kz + g \, [ \, \bQ \cdot \bA(s=0) - \bQ(s_n) \cdot \bA(s_n) \, ]
   =
   0 \>,
\end{equation}
and $s_n$ must be in the range $s_0 < s_n \le 0$.  This completes the solution using the method of characteristics. 

Since the Pauli term depends on the distribution function evaluated at $\kz = 0$, Eq.~\eqref{e:fsumsnI} needs to be solved explicitly for the case when $\kz = 0$.  For this case $s_n$ is a solution of the equation
\begin{equation}\label{e:snsolkzero}
   \bQ \cdot \bA(s=0)
   = 
   \bQ(s_n) \cdot \bA(s_n) \>.
\end{equation}
One such solution will always be $s_n=s=0$, or $t_n = t$.  Then using the relation $\Theta(0) = 1/2$, and solving \eqref{e:fsumsnI} for $f_0( \, t,\kperp,Q \, ) \defby f( \, t,0,\kperp,Q \, )$ gives
\begin{equation}\label{e:fk0sol}
   f_0( \, t,\kperp,Q \, )
   =
   \frac{ 
      S(t,\kperp,Q)/2 
      + 
      Z(t,\kperp,Q)
        }
        { 1 + S(t,\kperp,Q) } \>,
\end{equation}
where
\begin{equation}\label{e:Rdef}
   Z(t,\kperp,Q)
   =
   \sum_{s_n < 0} R(s_n,\kperp,Q) \>.
\end{equation}
This completes the solution of $f( \, t, \kperp, \kz, Q \, )$ using the method of characteristics.  The method used here for non-Abelian symmetries differs from that used for \QED\ since in this case we must numerically solve the trajectory equations for $\bQ(s)$ given the \emph{final} values of $x$, $k$.   Here the trajectory solution for $\bQ(s)$ does not have a conserved quantity as we found for $\kz(s)$ in Eq.~\eqref{e:pconstdef}.  Note that we do not need to have the values of $k$ or $\bQ$ at $s=s_0$ since by the backwards integration, we find the ``initial'' values needed to obtain the final values at $t$. 

However, it is not necessary to find the complete distribution function $f( \, t, \kperp, \kz, Q \, )$ in order to compute the currents since we can make use of the $\delta$-function in $C(t,\kperp,\kz,Q)$ to replace the integral over $\kz$ by an integral over $s'$.  This is a great advantage since it means that we only need to find the special distribution function $f_0( \, t,\kperp,Q \, )$.  
Substituting \eqref{e:BVintIsol} into Eqs.~\eqref{e:Jcontzdefs} and integrating over $\kz$, the convective color currents can be found from the equations,
\begin{subequations}\label{e:JcontzdefsII}
\begin{align}
   &\bJ_{\text{con}}^{t}(t)
   =
   \frac{g}{\pi^2} \int_0^{\infty} \!\!\!\kperp \rd\kperp\! 
   \int \rD Q \,
   \bQ \,
   \int_{s_0}^{0} \!\!\rd s' \, \kt(s') \,
   \label{e:JcontdefsII} \\
   & \> \times
   | \, g \, \bQ(s') \cdot \bE(s') \, | \,
   R[ \, s',\kperp,Q(s') \, ] / M \>, 
   \notag \\   
   &\bJ_{\text{con}}^{z}(s)
   =
   \frac{g}{\pi^2} \int_0^{\infty} \!\!\!\kperp \rd\kperp\!
   \int \rD Q \>
   \bQ \,
   \int_{s_0}^{0} \!\!\rd s' \, 
   \kt(s')  
   \label{e:JconzdefsII} \\
   & \> \times
   \frac{\kz(s',s,Q)}{\omega_{\kperp}(s',s,Q)} \,
   | \, g \, \bQ(s') \cdot \bE(s') \, | \,
   R[ \, s',\kperp,Q(s') \, ] / M \>, 
   \notag
\end{align}
\end{subequations}
where we have put
\begin{subequations}\label{e:kztomegatII}
\begin{align}
   \kz(s',s,Q)
   &=
   g \, 
   [ \, 
      \bQ(s') \cdot \bA(s') 
      - 
      \bQ \cdot \bA(s) \,
   ] 
   \label{e:kztII} \\
   \omega_{\kperp}(s',s,Q)
   &=
   \sqrt{ \kperp^2 + \kz^2(s',s,Q) + M^2 } \>.
   \label{e:omegatII}
\end{align}   
\end{subequations}

For the energy and pressures, substituting \eqref{e:BVintIsol} into Eqs.~\eqref{e:tensorcomps} gives
\begin{subequations}\label{e:tensorcompsII}
\begin{align}
   &\epsilon(t)
   =
   \frac{g}{\pi^2} \int_0^{\infty} \!\!\!\kperp \rd\kperp 
   \int \rD Q 
   \int_{s_0}^{0} \!\!\rd s' \, k_t(s') \,
   \label{e:energyII} \\ & \> \times
   \omega_{\kperp}(s',s,Q) 
   | \, g \, \bQ(s') \cdot \bE(s') \, | \,
   R[ \, s',\kperp,Q(s') \, ] / M \>, 
   \notag \\
   &\pperp(t)
   =
   \frac{g}{\pi^2} \int_0^{\infty} \!\!\!\kperp \rd\kperp 
   \int \rD Q 
   \int_{s_0}^{0} \!\!\rd s' \, k_t(s') \,
   \label{e:pperpII} \\ & \> \times
   \frac{\kperp^2}{\omega_{\kperp}(s',s,Q)} 
   | \, g \, \bQ(s') \cdot \bE(s') \, | \,
   R[ \, s',\kperp,Q(s') \, ] / M\>, 
   \notag \\
   &\pz(t)
   =
   \frac{g}{\pi^2} \int_0^{\infty} \!\!\!\kperp \rd\kperp 
   \int \rD Q 
   \int_{s_0}^{0} \!\!\rd s' \, k_t(s') \,
   \label{e:pzII} \\
   & \> \times
   \frac{\kz^2(s',s,Q)}{\omega_{\kperp}(s',s,Q)} \,
   | \, g \, \bQ(s') \cdot \bE(s') \, | \,
   R[ \, s',\kperp,Q(s') \, ] / M \>. 
   \notag   
\end{align}
\end{subequations}
In Eqs.~\eqref{e:JcontzdefsII} and \eqref{e:tensorcompsII}, we only need to find the distribution function $f_0(t,\kperp,Q)$ to obtain the currents, energy, and pressures.  However we will still need to back-integrate the trajectory equations to find $\bQ(s')$.
%
%
\begin{figure}[t!]
   \centering
   \includegraphics[width=\columnwidth]{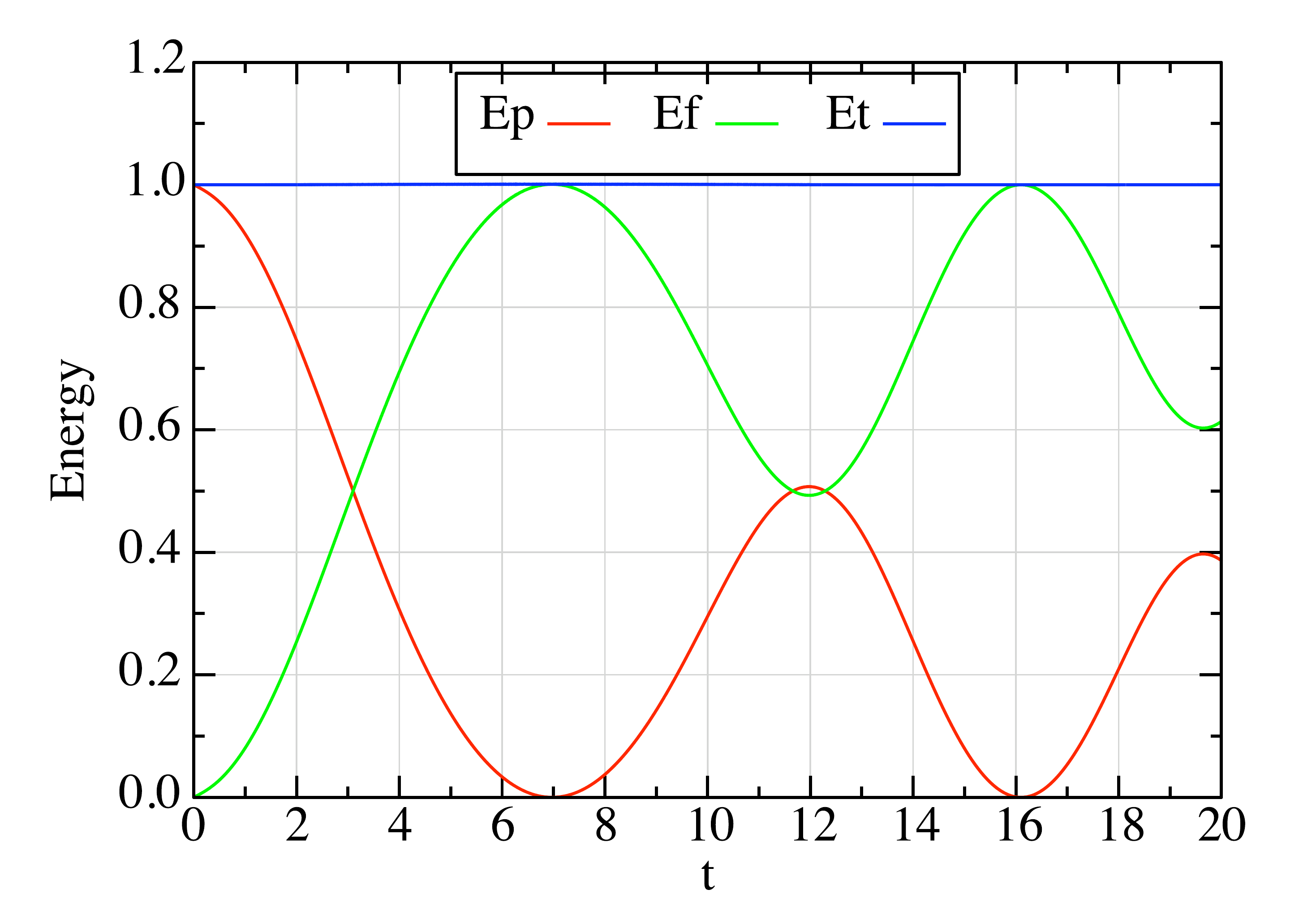}
   \caption{\label{f:Eng}
   Plot of the particle, field, and total energies as a function of $t$ 
   for a solution of the \BV\ equation
   with feedback for $SU(2)$ gauge fields.  Same initial conditions.}
\end{figure}
%
%
\begin{figure}[t!]
   \centering
   \includegraphics[width=\columnwidth]{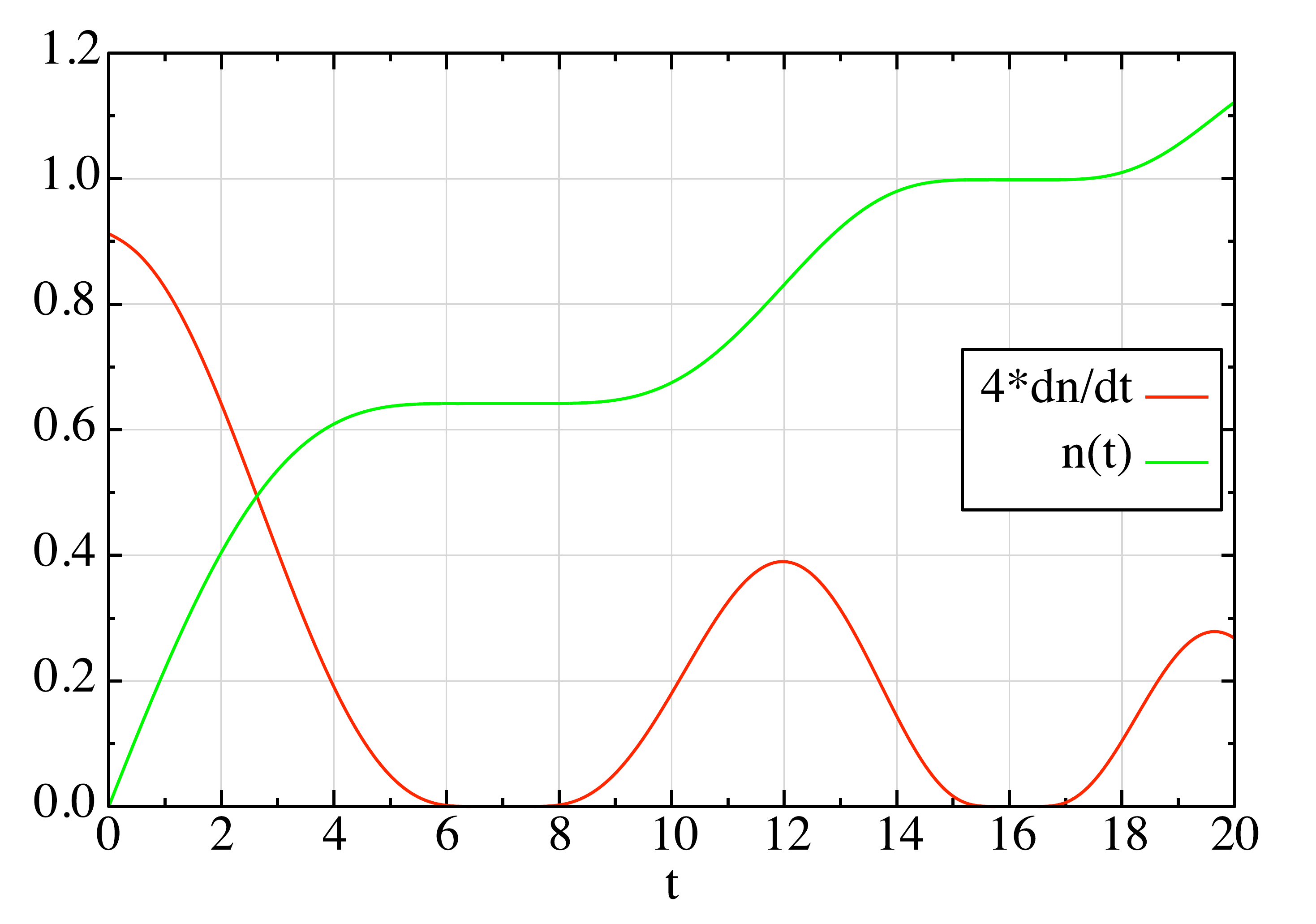}
   \caption{\label{f:RateNet}
   Plot of the particle production rate $\rd n / \rd t$ and total 
   particle production $n(t)$ from Eq.~\eqref{e:partprod} 
   as a function of $t$ for a solution of the \BV\ equation
   with feedback for $SU(2)$ gauge fields.  Same initial conditions.}
\end{figure}
%
%
\section{Polarization currents}
\label{s:polcurrents}

The total current is the sum of the convective and polarization currents,
\begin{equation}\label{e:JtotalJJ}
   \bJ^{\mu}(t)
   =
   \bJ^{\mu}_{\text{con}}(t)
   +
   \bJ^{\mu}_{\text{pol}}(t) \>,
\end{equation}
and satisfies the gauge field equation Eq.~\eqref{e:Jtcons},
\begin{equation}\label{e:JtconsIII}
   \partial_t \bJ^t(t)
   +
   g \, \bA(t) \times \bJ^z(t) 
   =
   0 \>.
\end{equation}
On the other hand, the convective part satisfies Eq.~\eqref{e:colorconsIII}.  So substituting \eqref{e:JtotalJJ} into \eqref{e:JtconsIII} and using \eqref{e:colorconsIII}, we find that the polarization current satisfies
\begin{equation}\label{e:polequ}
   \partial_t \bJ_{\text{pol}}^{t}(t)
   +
   g \, \bA(t) \times \bJ^{z}_{\text{pol}}(t)
   =
   - g \, \bV(t) \>,
\end{equation}
which fixes $\bJ_{\text{pol}}^{t}(t)$ in terms of $\bJ^{z}_{\text{pol}}(t)$ and $\bV(t)$.  Substituting \eqref{e:stermI} into Eq.~\eqref{e:Vdef}, $\bV(t)$ is given by
\begin{equation}\label{e:VdefII}
   \bV(t)
   =
   \frac{g}{\pi^2} \int_0^{\infty} \!\!\!\kperp \rd\kperp\!
   \int \rD Q \>
   \bQ \, | \, g \, \bQ \cdot \bE(t) \, | \, R(t,\kperp,Q) \>.
\end{equation}
The current component $\bJ^{z}_{\text{pol}}(t)$ is obtained from energy conservation.  Substituting \eqref{e:stermI} into Eq.~\eqref{e:GdefB} gives\begin{align}
   G^{t}(t)
   &=
   \frac{g}{\pi^2}
   \int_{0}^{\infty} \!\! \kperp \, \rd\kperp\! \int \rD Q
   \label{e:Gvalue} \\
   & \qquad\qquad \times
   \omega_{\kperp,0} \,
   | \, \bQ \cdot \bE(t) \, | \, 
   R(t,\kperp,Q) \>.
\end{align}
Now we can always write,
\begin{equation}\label{e:cando}
   | \, \bQ \cdot \bE(t) \, |
   =
   \Sgn{ \bQ \cdot \bE(t) } \,
   \bQ \cdot \bE(t) \>.
\end{equation}
So $G^{t}(t) = \bE(t) \cdot \bJ^{z}_{\text{pol}}(t)$, where
\begin{align}
   \bJ^{z}_{\text{pol}}(t)
   &=
   \frac{g}{\pi^2}
   \int_{0}^{\infty} \!\!\! \kperp \, \rd\kperp\! \int \rD Q
   \label{e:Jzpolzdef} \\
   & \qquad \times
   \omega_{\kperp,0} \, 
   \Sgn{ \bQ \cdot \bE(t) } \, \bQ \, 
   R(t,\kperp,Q) \>.
   \notag
\end{align}
So the rate of change of particle energy density is given by
\begin{equation}\label{e:depsds}
   \partial_t \, \epsilon(t)
   =
   \bE(t) \cdot \bJ^{z}(t) \>,
\end{equation}
where $\bJ^{z}(t)$ is now the total $z$-component of the current.  Adding the field  energy density from Eq.~\eqref{e:energya} to the particle density, we find that the total energy density is conserved:
\begin{equation}\label{e:energyiscons}
   \partial_t \,
   \bigl [ \,
      \epsilon(t) + E^2(t)/2 \,
   \bigr ]
   =
   0 \>.
\end{equation}
So in this section, we have found equations for the polarization contribution to the current by requiring that total energy, including field energy, be conserved for our source terms.  The polarization current is a required modification of the gauge field equations to account for the creation of particle and anti-particle pairs.

%
%
\section{Particle production}
\label{s:partprod}

Integrating the \BV\ equation \eqref{e:BVeqII} over $\kz$ and $\bQ$ gives
\begin{equation}\label{e:partprod}
   \frac{ \partial \, n( t, \kperp )}{\partial t}
   =
   \int \frac{\rD Q}{2\pi} \, 
   | \, g \, \bQ \cdot \bE(t) \, | \, 
   R(t,\kperp,Q) \>,
\end{equation}
where the particle density $n( t, \kperp )$ is given by
\begin{equation}\label{e:partden}
   n( t, \kperp )
   =
   \int_{-\infty}^{+\infty} \frac{\rd \kz}{2\pi}
   \int \rD Q \,
   f( t, \kperp, \kz, Q ) \>.
\end{equation}
A plot of the particle density $n( t, 0 )$ as a function of time for the case of \oneplusone\ dimensions is shown in Fig.~\ref{f:RateNet}.
%
%
\begin{figure}[t!]
   \centering
   \includegraphics[width=\columnwidth]{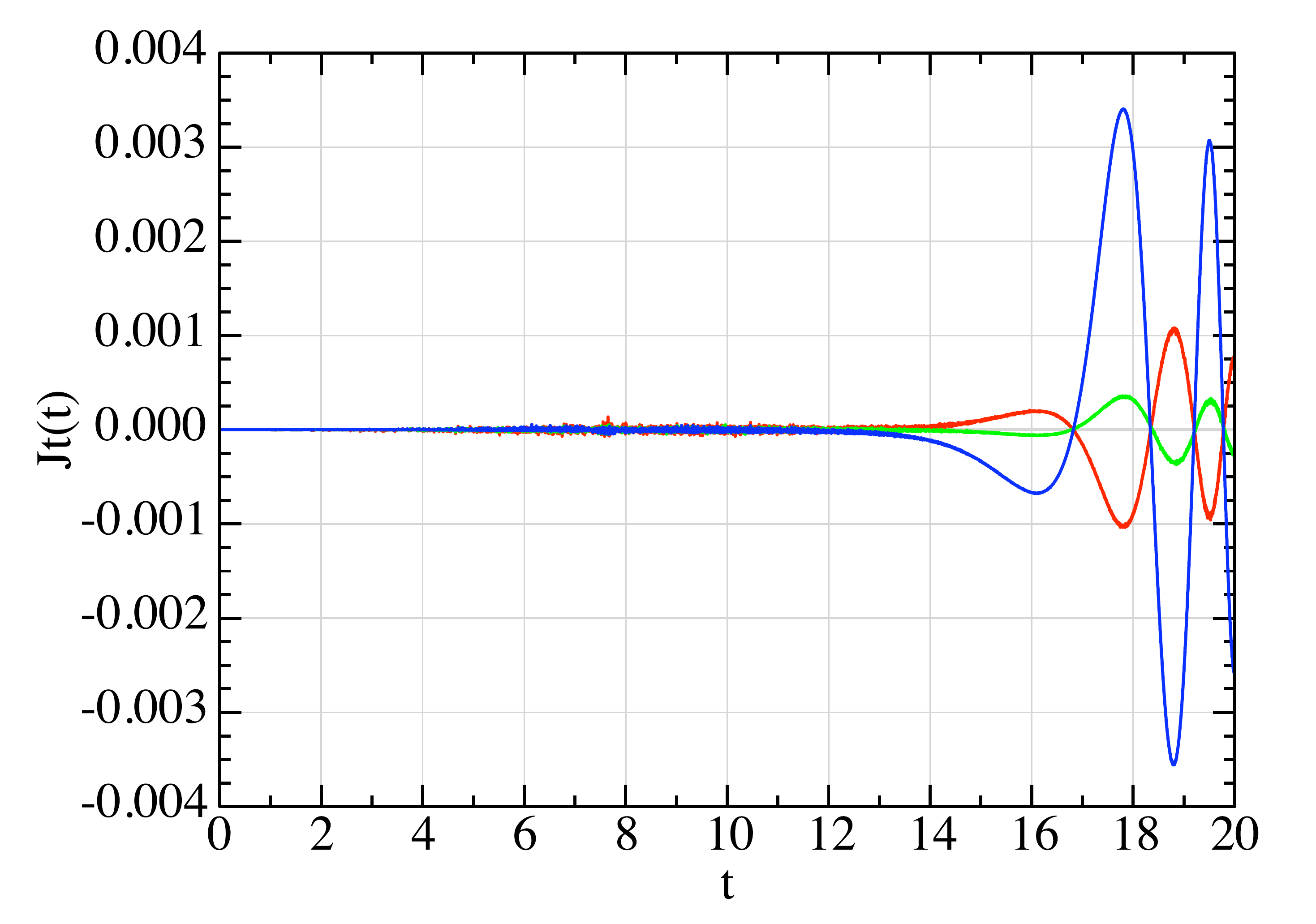}
   \caption{\label{f:Jct}
   Plot of the average value of $\bJ_{\text{con}}^{t}(t)$ as a function 
   of $t$ for a solution of the \BV\ equation
   with feedback for $SU(2)$ gauge fields.  Same initial conditions.}
\end{figure}
%
%
\begin{figure}[t!]
   \centering
   \includegraphics[width=\columnwidth]{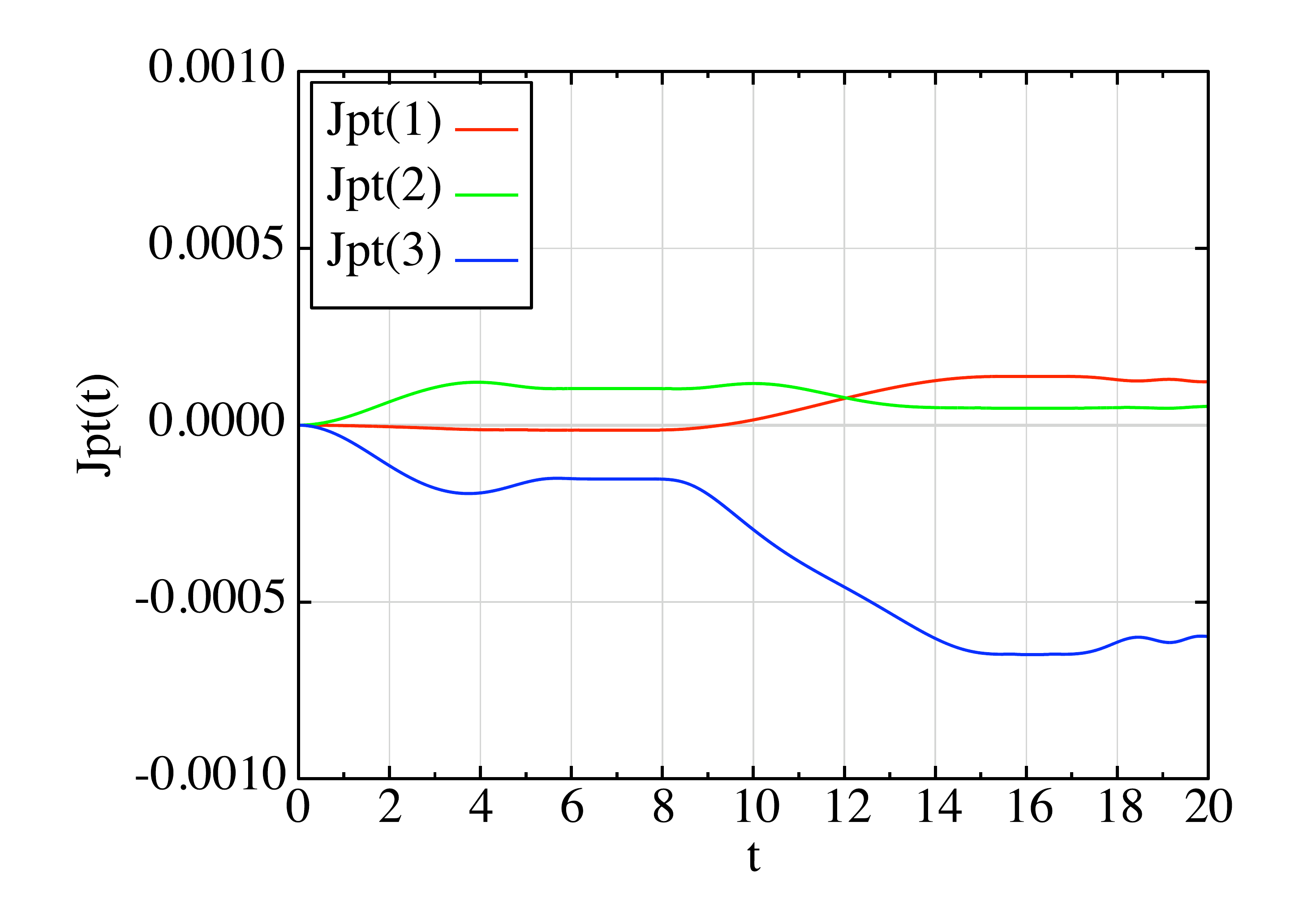}
   \caption{\label{f:Jpt}
   Plot of the average value of $\bJ_{\text{pol}}^{t}(t)$ as a function 
   of $t$ for a solution of the \BV\ equation
   with feedback for $SU(2)$ gauge fields.  Same initial conditions.}
\end{figure}
%
%

%
%
\section{Numerical methods}
\label{s:numerical}

Let us review the equations we need to solve.  The fields satisfy the equations
\begin{subequations}\label{nr.e:fieldsII}
\begin{align}
   \partial_t \, \bA(t)
   &=
   - 
   \bE(t) \>,
   \label{e:fieldsIIa} \\
   \partial_t \, \bE(t)
   &=
   -
   \bJ^{z}(t) \>,
   \label{e:fieldsIIb}
\end{align}
\end{subequations}
where the current is given as the sum of convective and polarization currents,
$\bJ^{\mu}(t) = \bJ_{\text{con}}^{\mu}(t) + \bJ_{\text{pol}}^{\mu}(t)$.  Components of the currents are given by Eqs.~\eqref{e:JcontzdefsII} and \eqref{e:Jzpolzdef}.
$\bJ_{\text{pol}}^{t}(t)$ is given by the solution of Eq.~\eqref{e:JtconsIII}, which is stepped out along with Eqs.~\eqref{nr.e:fieldsII}.  Back integrations of the trajectory equations of motion are required to find $\kt(s')$, $\kz(s')$, and $\bQ(s')$ for the currents.  The trajectory equations are given by
\begin{subequations}\label{nr.e:particleeom}
\begin{align}
   M \, \frac{\rd \kt(s)}{\rd s}
   &=
   g \, \bQ(s) \cdot \bE(s) \, \kz(s) \>,
   \label{nr.e:peoma} \\
   M \, \frac{\rd \kz(s)}{\rd s}
   &=
   g \, \bQ(s) \cdot \bE(s)  \, \kt(s) \>,
   \label{nr.e:peomb} \\
   M \, \frac{\rd \bQ(s)}{\rd s}
   &=
   g \, \bA(s) \times \bQ(s)  \, \kz(s) \>,
   \label{nr.e:peomc} 
\end{align}
\end{subequations}
with
\begin{equation}\label{nr.e:additional}
   M \, \frac{\rd t(s)}{\rd s}
   =
   \kt(s) \>,
   \qquad
   M \, \frac{\rd z(s)}{\rd s}
   =
   \kz(s) \>.   
\end{equation}
The backward integration is started at a proper time point $s = 0$ where $t(0) = t$.  At this point, particles are created with zero $z$-component of momentum, $\kz(t) = 0$.  So given values for $\kperp(0) = \kperp$ and $\bQ(0) = \bQ$, and using the parametric equations, the ``initial'' condition at $s=0$ for the backward integration is then specified as
\begin{alignat}{2}
   t(0) &= t \>,
   \qquad
   \kt(0) = \sqrt{\kperp^2 + M^2} \>, 
   \label{nr.e:intcond} \\
   z(0) &= 0 \>,
   \qquad
   \kz(0) = 0 \>.
   \notag
\end{alignat}
Eqs.~\eqref{nr.e:particleeom} and \eqref{nr.e:additional} are integrated backward to $t=0$, using the known values of $\bA(s')$ and $\bE(s')$ for $s_0 \le s' \le 0$, obtaining and values for $\kt(s')$, $\kz(s')$, and $\bQ(s')$.  Using these values, the integral over $s'$ for the currents, energy, and pressures can be done, and the process repeated for all values of $\kperp$ and $\bQ$ needed for the integral over $s'$.  During this back integration, it is also necessary to find the special distribution function $f_0(t,\kperp,Q)$ at each step.  This is given by Eqs.~\eqref{e:fk0sol} and \eqref{e:Rdef} and stored globally.  

A standard fourth order Runge-Kutta routine is used to step out the field equations  \eqref{nr.e:fieldsII} and the back integrations \eqref{nr.e:particleeom}.  In order to improve the accuracy of the Runge-Kutta method for the field equations, it is desirable to compute the derivative of the $z$-component of the current at each step, and use this for a linear interpolation for the currents during the steps.  Since the polarization current is generally quite small, we use only the derivative of the $z$-component of the convective current.  This is given by
\begin{align}
   &\partial_t \, \bJ^{z}_{\text{con}}(t)
   \label{nr.e:dJzconds} \\
   & =
   \frac{g}{\pi^2}
   \int_{0}^{\infty} \!\!\! \kperp \, \rd\kperp\! \int \rD Q \,
   ( \kperp^2 + M^2 ) \, \bQ 
   [ \, g \, \bQ \cdot \bE(t) \, ]
   \notag \\ & \quad \times
   \int_{s_0}^{0} \rd s' \, \frac{\kt(s')}{M} \, 
   \frac{ | \, \bQ(s') \cdot \bE(s') \, | \,
          R[ \, s',\kperp,Q(s') \, ] }
   { \omega_{\kperp}^3(s',s,Q) } \>.
   \notag
\end{align}

The color particle number density is obtained from $\bJ^{t}(t) = \bJ_{\text{con}}^{t}(t) + \bJ_{\text{pol}}^{t}(t)$.  The $t$-component of the convective current is obtained from Eq.~\eqref{e:JcontdefsII} and the $t$-component of the polarization current is obtained by the solution of Eq.~\eqref{e:JtconsIII}.

Some results are shown in Figs.~\ref{f:A}--\ref{f:RateNet} for the case where we have set $M=1$, $g=1$, and taken $\bA(0) = 0$ and $\bE(0) = ( \, 4, 2, 1 \, )$ and ignored the Pauli correction.  Here we took $\rd t = 0.002$ and $\rd s = 0.0005$ with $320$ values of $\bQ$ distributed in more or less equal area triangles about the surface of the $Q$-sphere.  This seems to be sufficient for values out to $t=20$.  The program took about 24 cpu hours on a desktop-type machine.

In Figs.~\ref{f:A} and \ref{f:E}, we see that the fields $\bA(t)$ and $\bE(t)$ oscillate with a period of approximately 20 in units of the mass $m$.  $\bA(t)$ remains negative during this time interval.  The $z$-component of the convective current $\bJ_{\text{con}}^{z}(t)$, shown in Fig.~\ref{f:Jcz}, appears to saturate even though there is no Paui correction factor here.  The polarization current, shown in Fig.~\ref{f:Jpz}, is about one-tenth the size of the convective current but is necessary to create plasma oscillations and is required for energy conservation.  In Fig.~\ref{f:Eng}, we plot the particle and field energies.  Energy is exchanged between these two energies with the total energy conserved to an accuracy of about 1\%.  The rate of particle production and total particle production is shown in Fig.~\ref{f:RateNet}.  Over half the total production is during the first few oscillations of the fields.  

The time-components of the convective current should vanish since the integrand in Eq.~\eqref{e:JcontdefsII} is odd under $\bQ \rightarrow -\bQ$.  As a check, we show by direct calculation in Fig.~\ref{f:Jct} that this component of current is less than $5 \times 10^{-3}$ in magnitude for $t < 20$.  The calculation begins to show instability for $t > 14$.  Likewise $\bV(t)$ vanishes since the integrand in Eq.~\eqref{e:VdefII} is odd under reversal of $\bQ$.  Eq.~\eqref{e:polequ} then reduces to
\begin{equation}\label{e:polequII}
   \partial_t \bJ_{\text{pol}}^{t}(t)
   +
   g \, \bA(t) \times \bJ^{z}_{\text{pol}}(t)
   =
   0 \>.
\end{equation}
Numerical solutions of \eqref{e:polequII} are shown in Fig.~\ref{f:Jpt}.  Here we see that all components of $\bJ_{\text{pol}}^{t}(t)$ are less that $1 \times 10^{-3}$ which, to the accuracy of this calculation, is consistent with zero.  So the auxiliary Eq.~\eqref{e:Jtcons} is satisfied by this calculation.

%
%
\section{Conclusions}
\label{s:conclusions}

In this paper, we have derived semi-classical transport equations for particle production by color gauge fields in a quark plasma with \SUtwo\ gauge symmetry.  We solved a particularly simple Cartesian \oneplusone-dimensional case, where the electric field is constrained to be in the $z$-direction, and showed how to obtain numerical solutions to the \BV\ equation with a Schwinger-type source term, starting with no particles and an initial large color field.  Our results conserve the average color and energy of the plasma, and satisfy the gauge field conditions.  

With additional computing power, our calculation can be extended to boost-invariant \threeplusone-dimensions for the \SUtwo\ case.  It would also seem possible to extend the calculation to the more interesting \SUthree\ case, where we could explore particle production as a function of the two conserved Casimir invariants.  

%
%

\begin{acknowledgments}
This work was performed in part under the auspices of the United States Department of Energy. The authors would like to thank the Santa Fe Institute for its hospitality during the completion of this work.  One of us (JFD) would like to thank Robert Carrier of the University of New Hampshire Research Computing Center for helpful discussions.  He would also like to thank Matthew Minuti for computing advice and Silas Beane for use of computing resources. 
\end{acknowledgments}


\bibliography{johns}
%
%
\end{document}